# Improved Performance of Organic Light-Emitting Transistors Enabled by Polyurethane Gate Dielectric

*Arthur R. J. Barreto[a], Graziâni Candiotto[b], Harold J. C. Avila[c], Rafael S. Carvalho[a], Aline Magalhães dos Santos[a], Mario Prosa[d], Emilia Benvenuti[d], Salvatore Moschetto[d], Stefano Toffanin[d], Rodrigo B. Capaz[b,e], Michele Muccini[d], Marco Cremona[a]\**

[a] Organic and Molecular Optoelectronics Laboratory (LOEM), Department of Physics, Pontifical Catholic University of Rio de Janeiro, Rio de Janeiro – RJ, 22451-900, Brazil

[b] Institute of Physics, Federal University of Rio de Janeiro, Rio de Janeiro - RJ, 21941-909, Brazil.

[c] Department of Physics, University of Atlantic, Puerto Colombia, Atlántico, Colombia

[d] Institute of Nanostructured Materials (ISMN), National Research Council (CNR), via P. Gobetti 101, 40129 Bologna, Italy

[e] Brazilian Nanotechnology National Laboratory (LNNano), Brazilian Center for Research in Energy and Materials (CNPEM), Campinas, SP 13083-100, Brazil.

*cremona@fis.puc-rio.br

ABSTRACT

Organic light-emitting transistors (OLETs) are multifunctional optoelectronic devices that combine in a single structure the advantages of organic light emitting diodes (OLEDs) and organic field-effect transistors (OFETs). However, low charge mobility and high threshold voltage are critical hurdles to practical OLETs implementation. This work reports on the improvements obtained by using polyurethane films as dielectric layer material in place of the standard poly(methylmethacrylate) (PMMA) in OLET devices. It was found that polyurethane drastically reduces the number of traps in the device thereby improving electrical and optoelectronic device parameters. In addition, a model was developed to rationalize an anomalous behavior at the pinch-off voltage. Our findings represent a step forward to overcome the limiting factors of OLETs that prevent their use in commercial electronics by providing a simple route for low-bias device operation.

## I.     INTRODUCTION

Organic light-emitting transistors (OLETs) are multifunctional optoelectronic devices that combine in a single device the advantages of light emission of an organic light emitting diode (OLED) and the electrical switching characteristics of an organic field-effect transistor (OFET)[1,2]. Due to these functionalities, OLETs offer the opportunity to simplify circuit design and considerably increase the number of new applications of organic semiconductor-based devices, including the highly desirable flexible integrated optoelectronics, fully organic active matrix displays, electrically driven lasers, nanoscaled light-sources etc[3,4,5,6,7]. In a simplified picture, the work and operation of a unipolar OLET can be viewed as a traditional OFET in its electrical characteristics, with a conduction channel formed by majority charge carriers. In addition, depending on the applied bias, the injection of the minority charge carrier can occur, promoting radiative recombination at the proximity of the drain electrode, resulting in a controlled emission of light.

Two of the greatest challenges in the development of all organic-based integrated circuits are the relatively low mobility and the high threshold voltage characteristics of OFETs and OLETs when compared to traditional silicon-based electronics. In the last years, considerable efforts have been made to increase charge-carrier mobility and decrease operating voltages aiming at decreasing power consumption and improving operational stability. In OFETs, and consequently in OLETs, charge mobility greatly depends on the degree of molecular (structural) and energetic order, the latter being critically affected by semiconductor/dielectric interface morphology. In order to minimize potential energy barriers and facilitate charge carrier transport low levels of molecular and energetic disorder are required[8,9,10].

Several dielectric materials were thoroughly characterized in OLET devices[11,12,13], resulting in both $SiO_2$ and poly(methyl methacrylate) (PMMA) as benchmark dielectrics, with k = 3.9 and k = 3.0 to 4.5, respectively[14,15,16]. The use of these dielectric materials results in high operating voltages, *i.e.*, more than 20 V threshold voltage and operating voltage in the range of 80 to 100 V. Several approaches have been explored to allow for low-voltage operation, from vertical geometry OLETs[17] which, unfortunately, present unstable OFF state and high leakage current density, to passivation techniques[18], through employment of high-k dielectric layers[19]. In fact, the most common approach to this problem has been to use high-k metal oxides with a large capacitance density such as $Al_2O_3$, $HfO_2$, $TiO_2$, and $ZrO_2$[20,21,22,23]. However, the interface between organic semiconductors and high-k dielectrics is difficult to be controlled and optimized. Indeed, it typically contains deep traps due to dipolar chemical groups and hydroxyl groups (-OH) at the interface. They both act as localized defects that trap charge carrier, lower their mobility and hinder the formation of the charge transport channel, thereby degrading the electrical behavior of the device in terms of operational stability, low mobility and undesired hysteresis[24,25,26].

The detrimental high polarization at the dielectric/organic semiconductor interface can be avoided by using low-k dielectrics, as shown by previous works[16] and recently by our group[27] in which polyurethane (PU) was used as gate dielectrics (k = 3.0) combined with regioregular poly(3-hexylthiophene) (rrP3HT) in a bottom-contact top-gate OFET. As these works demonstrate, depending on the chemical nature of the dielectric, it can induce a different behavior from that seen in the benchmark (PMMA). In addition, PU has also been used as dielectric layer in OFETs [15,28,29] and as substrates for OFETs[30], AMOLED display panels[31] and several stretchable devices[32].

In this work, we report the fabrication of an OLET using PU (PU-based OLET) as dielectric layer and 2,7-dioctyl[1]benzothienol[3,2-b][1]benzothiophene (C8-BTBT) as small molecule organic semiconductor combined with an emission layer of tris(4-carbazoyl-9-ylphenyl)amine (TCTA) doped with fac-tris[2-phenylpyridinato-C2,N]iridium(III) (Ir(ppy)₃). Our results show that the PU dielectric layer enhances the device's performance and decreases operating voltages in comparison to the benchmark device based on PMMA. The measured electrical and optical properties of PU-based OLET shows great improvement when compared to PMMA-based OLET, with lower threshold voltage and subthreshold swing together with higher mobility and optical power output. Also, our analysis shows that changing the dielectric layer greatly decreases the number of traps at the semiconductor/dielectric interface. A model is also presented to better understand an anomalous behavior (lump) that is observed in the output characteristic curves near the pinch-off voltage.

## II.     MATERIALS AND METHODS

Polyurethane was used as dielectric layers to make bottom-gate/top-contact devices (Figure 1A). Both were spin-coated on clean patterned glass/ITO substrates (Lumtec, 15Ω/sq). PU (Ellastolan® PU 85A10 from BASF) (Figure 1D) is made up of block copolymers with alternating hard segment (based on 4,4 methylene diphenyl diisocyanate) and soft segment (based on and 1,4 butane diol) without the addition of plasticizers. PU was dissolved in 1,4-dioxane and spin-coated to have 700 nm as final thickness, following the procedure in[27]. PMMA was used as reference dielectric and all comparison data are consistent with the literature[33].

The semiconducting layers and Ag electrodes were deposited via thermal evaporation (base pressure $10^{-6}$ torr). The rigid-band energy diagram is shown in Figure 1B. First, a 35 nm-thick C8-BTBT (Sigma-Aldrich) layer was deposited at 0.1 Å/s rate, then the 7% wt doped 60 nm-thick host-guest emissive layer was deposited using Ir(ppy)$_3$ (Sigma Aldrich) as guest and TCTA (Sigma Aldrich) as host. Finally, 70 nm-thick Ag electrodes were deposited (0.5 Å/$s$) using interdigitated shadow masks (L = 70 µm and W = 84 mm) (Figure 1C). The molecular structures of the different materials are shown in Figure 1D.

The devices were encapsulated inside a nitrogen-filled glove-box with glass lids and epoxy glue. Once encapsulated, the electrical measurements were performed in ambient conditions using a probe station from Cascade Microtech model EPS150X together with a Keysight B2912A semiconductor analyzer using its own software for data acquisition. The light output was obtained using a Newport 818-UV photodiode in direct contact with the glass substrate.

The EQE was measured by collecting the emitted photons using the same calibrated silicon photodiode (Newport 818-UV) together with its known responsivity. Although this measurement could underestimate the total power[34], it accurately measures the relative efficiency between devices. EQE was then calculated by the photodiode current combined with photodiode spectral responsivity, source-drain currents and emission spectra of the devices as per standard literature methods[35].

The photoluminescence (PL) and the electroluminescence (EL) spectra were obtained using a spectrofluorometer from Photon Technology International (PTI) Quanta Master model 40. The thickness of each layer was measured by profilometer Veeco Dektak 150 and Atomic Force Microscopy (AFM) measurements were performed using a NT-MDT Old Smena. Images were taken in semi-contact mode with a 10 nm radius tip at a resonance of 330 MHz.

Density functional theory (DFT) was used to investigate the molecular properties of the organic compounds at atomic level, such as dipole moment. Calculations were performed using the hybrid functional B3LYP (Becke three-parameter Lee-Yang- Parr)[36] along with a split-valence double-zeta polarized based in Gaussian type orbitals (6-31G (d, p))[37,38] present in Gaussian 03 package.

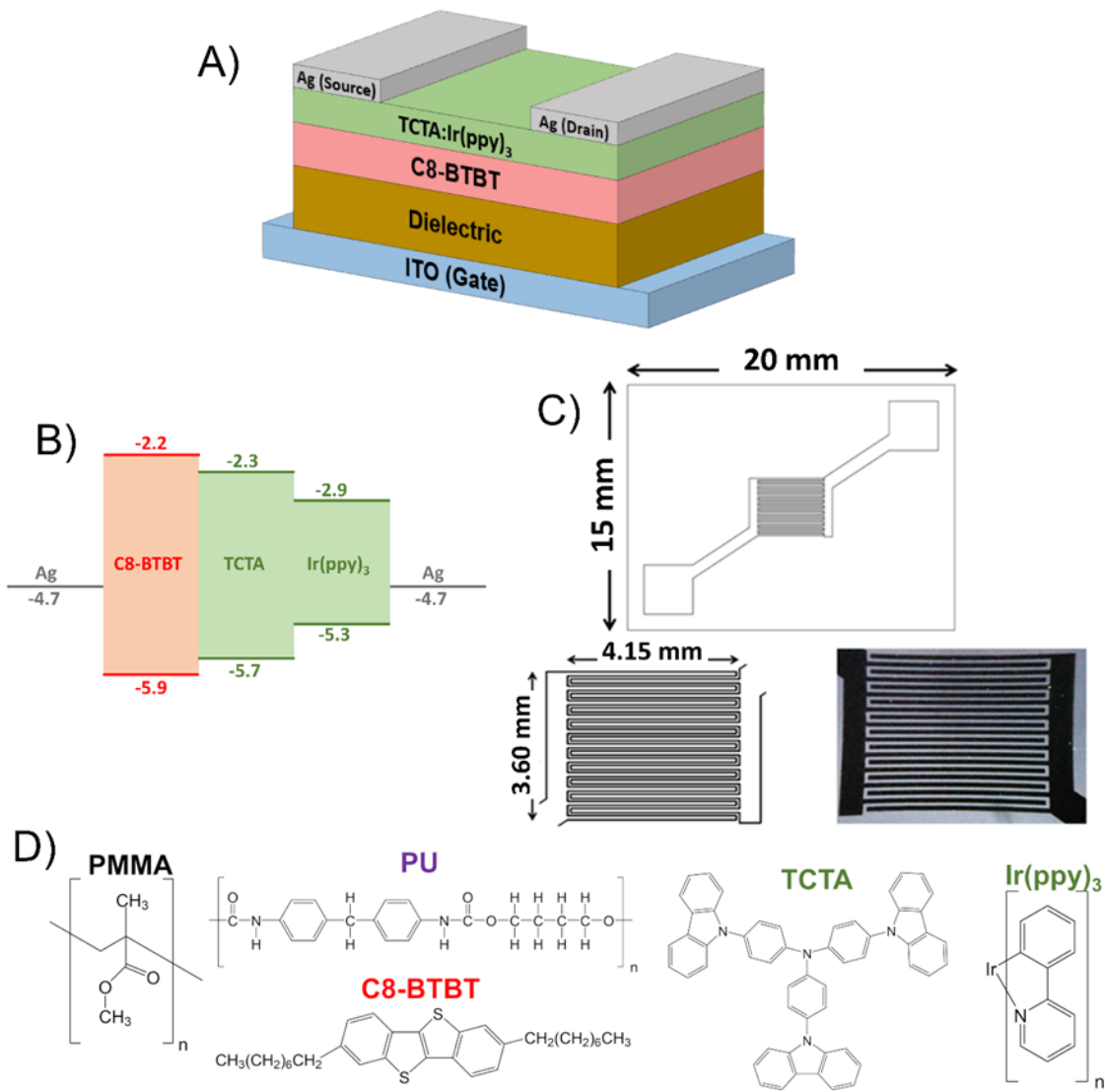

**Figure 1**. A) Schematic illustration of OLET architecture used, B) energy level diagram, C) interdigitated mask and D) molecular structure of PMMA, PU, C8-BTBT, TCTA and Ir(ppy)₃.

## III. RESULTS AND DISCUSSION

### a. Morphological Analysis

For the development of high quality OLETs, the solid-state organization and the surface quality of each layer of the device are of pivotal importance to ensure an efficient OLET operation providing state-of-the art performance. Accordingly, morphological analyses were carried out to investigate the surface of the PU dielectric layer and its influence on the solid-state organization of the overlying layers.

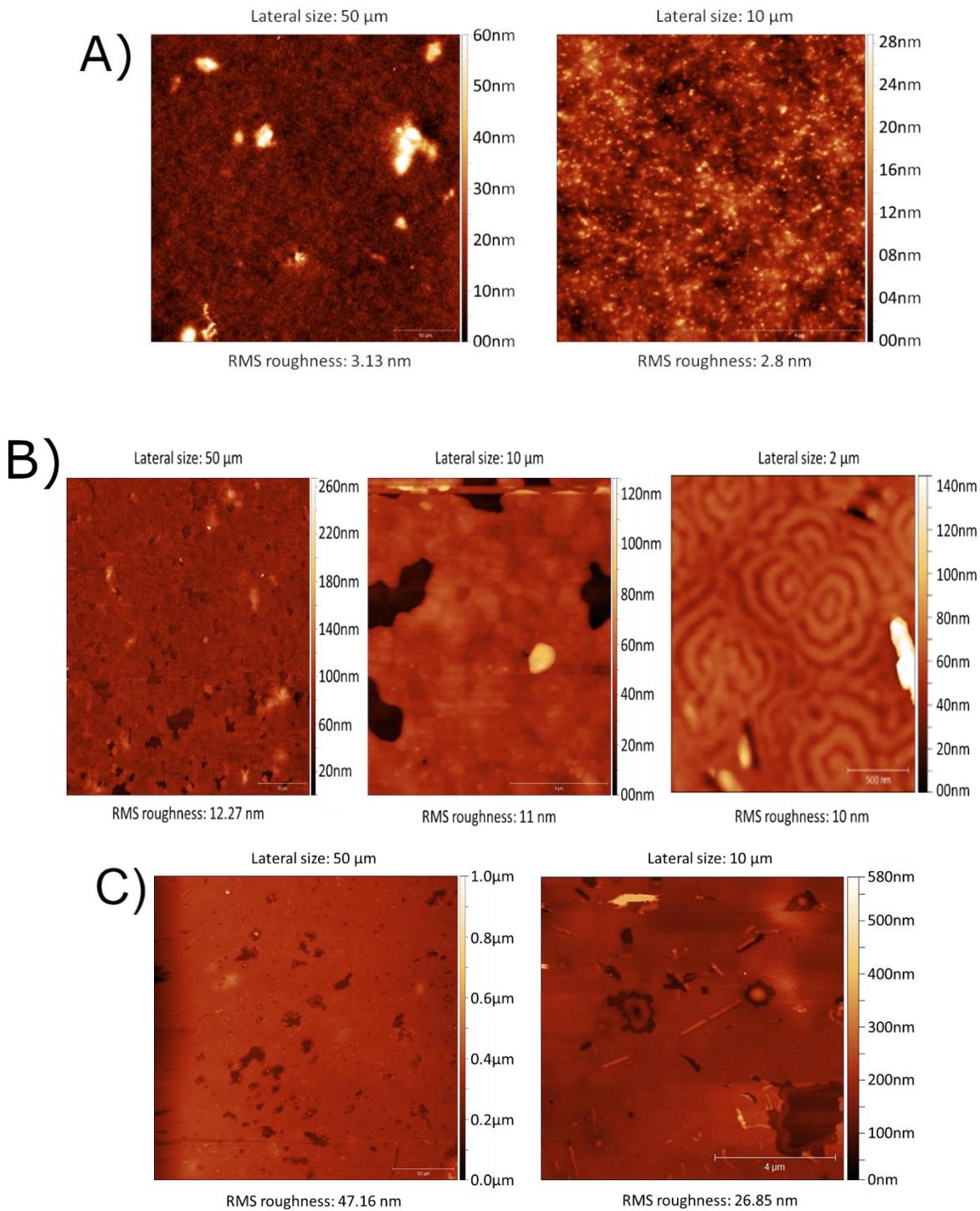

**Figure 2**. AFM images for A) PU, grown on glass, B) C8-BTBT, grown on PU and C) TCTA:Ir(ppy)₃, grown on C8-BTBT.

PU's surface was analyzed through AFM and the morphology is shown in Figure 2A. The AFM images do not show significant cracks or pin-holes over different areas, in agreement with the literature[15]. The surface morphology for the semiconducting layer C8-BTBT and for the emission layer TCTA:Ir(ppy)$_3$ were also analyzed, and the morphologies are shown in Figure 2B and Figure 2C, respectively.

In particular, C8-BTBT shows the expected layer-by-layer growth modality[19,39] when deposited on top of PU. Although an increase in surface roughness and surface defects can be observed after the deposition of C8-BTBT and TCTA:Ir(ppy)$_3$ onto the PU layer, negligible influences on the quality of both semiconductor and emissive layers are expected, as highlighted by the state-of-the-art characteristics of the corresponding OLETs (vide infra).

### b. Device's characterization

Figure 3A shows the behavior of drain-source current ($I_{DS}$) in locus mode for the PU-based OLETs, where both drain-source voltage ($V_{DS}$) and gate-source voltage ($V_{GS}$) are swept at the same time, keeping $V_{DS} = V_{GS}$. This type of characterization ensures saturation regime throughout the entire sweep, maintaining the pinch-off near the drain electrode[40].

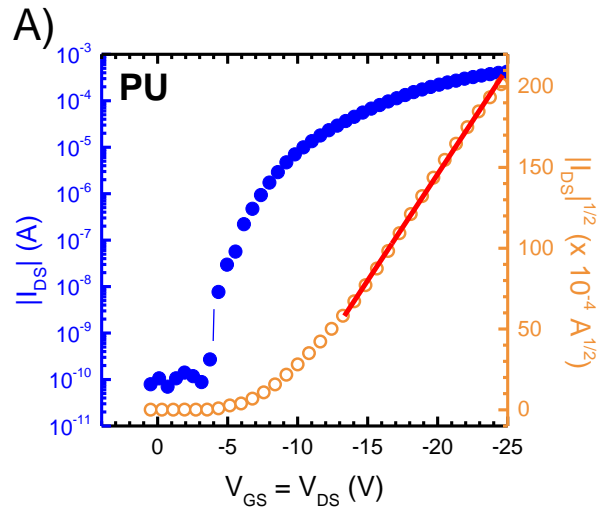

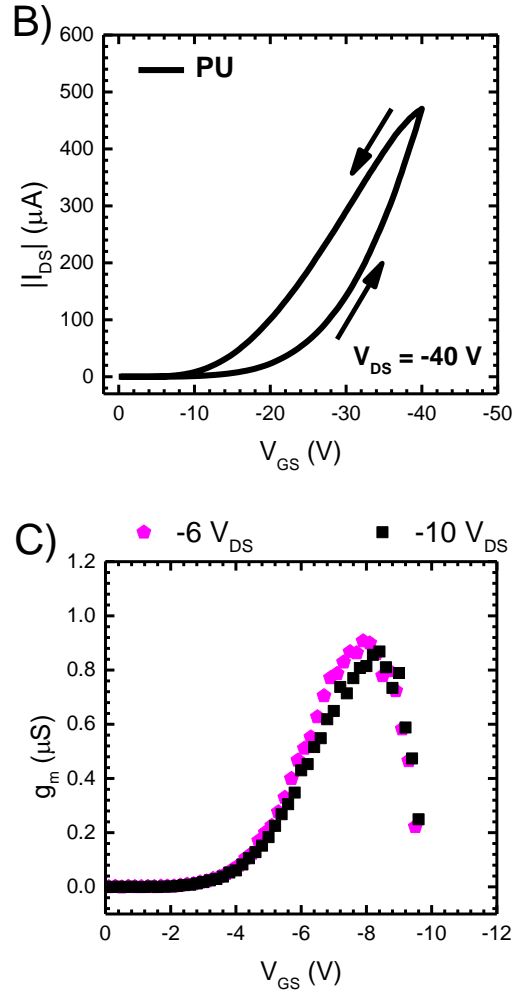

**Figure 3**. A) Locus characteristics (filled symbols) and linear fit (open symbols) for PU-based OLETs. B) Cyclic transfer characteristics curve for PU-based OLETs, black arrows indicate the sweep voltage direction. C) Transconductance curve for PU-based OLET.

Experimental parameters and the results of this analysis are summarized in Table 1. If compared to PMMA-based devices, the use of PU as a dielectric material shows a significant improvement of the OLET characteristics.

Specifically, a five-times decrease of threshold voltage is observed. Moreover, the emitted optical power and ON/OFF ratio are increased while the subthreshold swing (SS) value is reduced more than four times with a consequent reduction over three times in the density of interfacial traps (see electrical characterization). Interestingly, mobility and external quantum efficiency (EQE%) are doubled.

**Table 1**. Optoelectronic parameters of PU-based OLETS and PMMA-based OLETs. All electrical measurements represent an average over ten performing devices. *Values taken from[33].

| | PU | PMMA |
|---|---|---|
| $C_D$ ($\mu F/m^2$) | $51 \pm 2$[27] | $40 \pm 2$[27] |
| m (D/Å) | 0.178 | 0.442 |
| $V_{th}$ (V) | - $6.5 \pm 0.3$ | - 34* |
| SS (V/dec) | $-0.84 \pm 0.02$ | $-3.90 \pm 0.05$* |
| $N^1$trap ($\times 10^{11} cm^{-2}$) | 2.55 | 8.50* |
| $N^2$trap ($\times 10^{11} cm^{-2}$) | 4.21 | 16.20* |
| $\mu$ ($cm^2/Vs$) | $0.4 \pm 0.2$ | 0.19* |
| Optical power at -70V ($\mu W$) | $5 \pm 1$ | 0.15* |
| ON/OFF | $>10^6$ | $>10^6$* |
| EQE (%) | 0.04 ($V_{DS} = V_{GS} = -60V$) | 0.02 ($V_{DS} = V_{GS} = -100V$) * |

The optimization of the threshold voltage parameter is extremely important for a device to work at low voltages, since it provides the minimum $V_{GS}$ needed to create a conducting channel between the source and drain terminal. Another important parameter that must be optimized for the device to have better performance at low working voltages is subthreshold swing (SS). The SS parameter measures how fast the device change from OFF to ON state and it can be extracted from locus electrical curve using the relation SS = $dV_{GS}/d(\log|I_{DS}|)$. In literature, the significant improvements in $V_{th}$ and SS measured can be attributed to a decrease in charge density at the dielectric/semiconductor interface[41]. In this work, two methods were used to estimate the density of traps in the devices. 1) In the first method, it is assumed that the charges are accumulated and trapped close the dielectric/semiconductor interface of the OLET, thus the value of $V_{th}$ is equivalent to trapped charge on one side of an equivalent capacitor[42,43]. 2) In the second method, it is estimated via Subthreshold Swing. These two methods can be written as below.

Method 1
$$N_{trap}^1 \approx \frac{C_D}{q} V_{th} \qquad (1)$$

Method 2
$$N_{trap}^2 \approx \frac{C_D}{q} \left[ \frac{qSS}{k_b T \ln 10} - 1 \right] \qquad (2)$$

where q is the elementary electronic charge, $k_b$ is the Boltzmann's constant and T is the room temperature. Using methods 1 and 2, it was estimated for PU-based OLETs an interfacial trap density equal to $N_{trap}^1 \approx 2.6$ x $10^{11}$ cm$^{-2}$ and $N_{trap}^2 \approx 4.2$ x $10^{11}$ cm$^{-2}$. Comparing these values with their equivalents for the PMMA-based OLET ($N_{trap}^1 \approx 8.5$ x $10^{11}$ cm$^{-2}$ and $N_{trap}^2 \approx 16.2$ x $10^{11}$ cm$^{-2}$) from[33], the interfacial traps density in PU-based OLET is more than three times smaller than PMMA-based OLET.

Looking at the cyclic transfer characteristics curves for PU-based (Figure 3B) and PMMA-based OLETs (Figure 2a from[33]), in both cases it is observed a reversible hysteresis during the sweep voltage. The presence of hysteresis in the characteristic curves of organic devices is usually attributed to charge traps in the system, as shown by Hwang *et al.*[40]. The hysteresis exhibited by PU and PMMA-based devices can origin from different sources. PU-based device shows an anticlockwise hysteresis, where the forward sweep current is lower than backward sweep current. This behavior is observed in organic devices with relatively thick gate dielectrics in which there are chemical species that are capable of slow polarization (*e.g.* polar functionalities, ionic impurities and water molecules)[44]. Diversely, the PMMA-based device shows a clockwise hysteresis, where the forward sweep current is higher than backward sweep current. This behavior is characterized by structural defects of semiconductor films, dielectric surface functionalities and adsorbed small molecules (*e.g.,* oxygen) at the interface. These factors are able to generate long-lived charge traps present at the dielectric/semiconductor interface, where trapping and detrapping of holes and electrons take place[44].

The determination of OLET/OFET's parameters derived from anticlockwise hysteresis curves can be influenced by the effect of slow polarization dielectrics. This polarization can cause an increase in the accumulated charge in the semiconductor channel and, consequently, resulting in overestimated parameters. Therefore, the parameters were extracted from the very first forward sweep, in order to minimize the time exposure of the devices to the applied voltages and thus minimize the effects of polarization. This approach ensures that all parameters are a lower bound for these devices, *i.e.*, the accumulated charges due to polarization effects has the least possible influence on the parameters reported in this work.

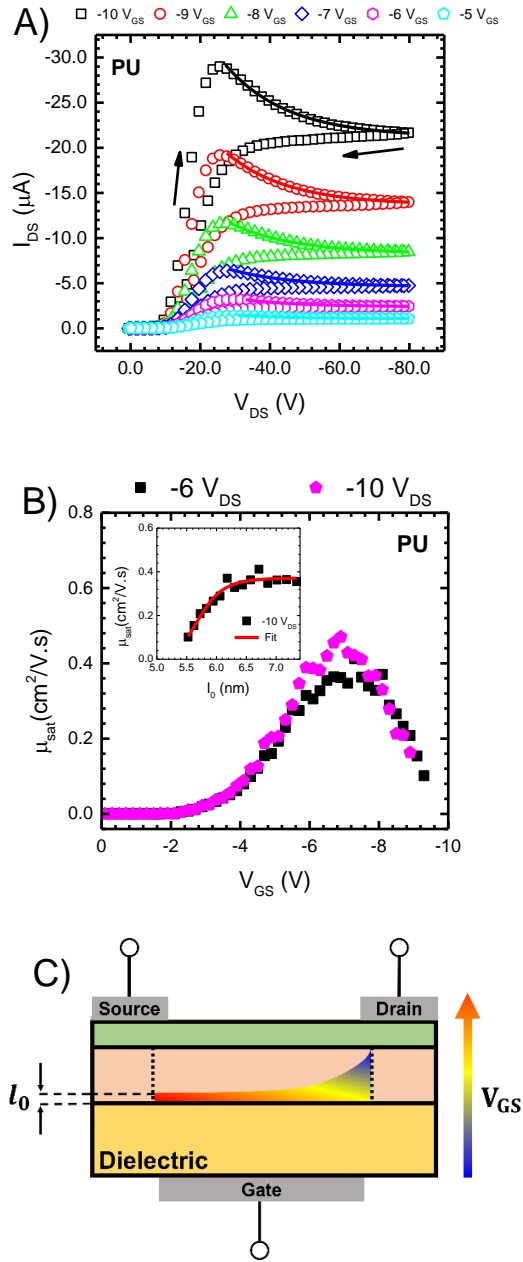

**Figure 4**. A) Output characteristic curves for PU-based OLETs with black arrows indicating the sweep voltage direction. Open symbols represent experimental data and solid lines represent the fitting using Eq. 8. B) Mobility curves as a function of $V_{GS}$ for PU. The inset shows the behavior

of $\mu_{sat}$ as a function of the channel bottleneck thickness $l_0$. C) Schematic representation for $l_0$ as a function of $V_{GS}$.

Some applications that mainly require amplification of small signals can benefit from the maximization of parameters like the transconductance $g_m = dI_{DS}/dV_{GS}$ and the output resistance $R_0 = g_m^{-1}$. Figure 3C shows the behavior of $g_m$ in different $V_{DS}$ for PU OLETs. As expected, the transconductance is independent of $V_{DS}$. Comparing the cases, it is observed that the maximum $g_m$ ~0.8 µS for the PU-based OLET is obtained around $V_{GS}$ ~ -8 V, while the same $g_m$ for PMMA devices is obtained under a much higher $V_{GS}$, around -47 V[33]. Thus, showing a considerable improvement in $g_m$ due to the use of PU as gate dielectric.

Figure 4A shows output characteristic curves for PU-based OLETs. Linear and saturation regimes are fully achieved as early as $V_{GS}$ ~ -3 V. It is observed that PU-based OLET reaches saturation currents above 20 µA under $V_{GS}$ = -10 V condition, while PMMA-based OLETs are not expected to show typical output curve behavior, since the conduction channel is not formed yet ($V_{th}$ = -34 V). PMMA-based OLET only reaches the same magnitude of current over $V_{GS}$ = -40V. Thus, the use of PU as a dielectric layer shows a relevant improvement in the reduction of OLET's working voltages.

Additionally, analyzing the output characteristic curves ($I_{DS}$ vs $V_{DS}$) for PU-based OLETs (Figure 4A), two distinct behaviors are observed: 1) the output characteristics shows a high $I_{DS}$ response for low values of $V_{GS}$ and 2) it shows a lump when approaching the pinch-off voltage. Such overall increase in $I_{DS}$ is associated with the reduction in threshold voltage and the increase in charge carrier's mobility in the system, *i.e.*, in the C8-BTBT/PU (semiconductor/dielectric) interface.

As saturation mobility ($\mu_{sat}$) plays a key role in OLET performance, its understanding is crucial[45]. In our study we determine saturation mobility $\mu_{sat}$ via linear best-fit using[43]:

$$I_{DS} = \frac{W}{2L} C_I \mu_{sat}(V_{GS} - V_{Th})^2, \qquad\qquad 3$$

and over the entire gate bias range using

$$\mu_{sat} = \left(\frac{d|I_{DS}|^{\frac{1}{2}}}{dV_{GS}}\right)^2 \frac{2L}{WC_I}, \qquad\qquad 4$$

where W and L correspond to the channel width and length of the OLETs, respectively. Figure 4B report the $\mu_{sat}$ for PU-based OLETs as a function of $V_{GS}$. This result shows that $\mu_{sat}$ is independent on $V_{DS}$ and the maximum $\mu_{sat} \sim 0.4$ cm$^2$/Vs is obtained for a $V_{GS}$ value around -7.5 V. On the contrary, the reference PMMA devices show a $\mu_{sat} \sim 0.2$ cm$^2$/Vs, that is two times lower.

The high mobility obtained for PU-based OLETs at low working $V_{GS}$ is associated with two factors: 1) the hydrophobic soft segment (CH$_2$) in the PU molecular structure is responsible for enabling strong molecule-molecule interaction, thus improving molecular ordering significantly at dielectric/semiconductor interface[7,15]. 2) The low dipole moment per unit length (m/L $\sim 0.178$ D/Å) of PU, as shown in[2,17] is responsible of the decrease of the random oriented dipoles at the dielectric/semiconductor interface and, as a consequence, of the surface energy decrease. The two factors mentioned above are frequently reported in the literature[7,17,18,19] as crucial to obtain higher mobility and $V_{th}$ stability.

Though, the presence of a lump in PU-based OLET output characteristic is a common feature especially observed in devices that have dipolar centers[32,46,47,48,49,50,51,52]. This behavior has been modeled considering a charge carrier mobility gradient in the direction perpendicular to the semiconductor/dielectric interface by Sworakowski *et al.*[45,53]. According to the Sworakowski model, charge traps at the semiconductor/dielectric interface can cause spatially inhomogeneous charge distribution in the organic semiconductor and a variation of charge mobility as a function of the distance normal to the semiconductor/dielectric interface. In this way, the mobility will depend on the electrostatic interaction energy, $E(l_0)$, generated by the charge traps in the conduction channel as a function of channel bottleneck thickness ($l_0$).

$$\mu(l_0) = \mu_0\left(1 + \zeta \exp\left(\frac{E(l_0)}{k_bT}\right)\right)^{-1}, \qquad\qquad 5$$

where $\mu_0$ is the mobility in a trap-free sample, $\zeta$ is a parameter that indicates the fraction of molecules that form local states. According to Sworakowski, for dipolar interactions, $E(l_0)$ scales as:

$$E(l_0) = \frac{A}{l_0^2}, \qquad\qquad 6$$

where A is proportional to the average dipole moment at the interface. The minimum effective channel thickness used in Sworakowski model can be independently estimated through the drift-diffusion equation, as shown by Seidel *et al.*[54]:

$$l_0 = \frac{4\varepsilon_s k_b T}{qC_I \left(V_{GS} - V_{th}\right)}, \qquad\qquad 7$$

where $\varepsilon_s$ is the semiconductor's permittivity.

Equation 7 shows that the effective channel thickness where charge carriers cross the device is inversely proportional to $V_{GS}$, so large values of $V_{GS}$ implies a reduction of the effective transport space in the semiconductor (bottleneck shown in Figure 4C) and, therefore, charge carriers will pass closer to the dielectric/semiconductor interface. This hinders the charge transport in the device, causing an increase in the channel resistance with a consequent decrease in $g_m$. The inset in Figure 4C shows the behavior of $\mu_{sat}$ as a function of $l_0$, where Eqs. (5-7) are used to fit the data (red line). A small value of $l_0$ means that the charge carries will cross the device close to the dielectric/semiconductor interface. The decrease in mobility observed when $l_0 \rightarrow 0$ is associated with an increase in the dipolar interaction (Eq. 6) that can be interpreted as the increase in the cross section between charge carrier and dipolar groups at the interface[45]. According to Veres *et al.*[16], the increase in interaction energy close to the interface is a determining factor for the mobility reduction observed in the Figure 4B. Thus, the increase in interaction energy can be interpreted as the emergence of a hindering current formed by scattering centers[55].

Therefore, our results allow us to conclude that the formation of lumps near the pinch-off regime is associated with the presence of slow polarization centers in gate dielectrics. Hence, while the linear regime maintains its well-known behavior, the presence of charge traps together with the bottleneck effect reduces the mobility in the device, so the current after the pinch-off regime can be written as

$$I_{DS,sat} = I_{pinch} \left[1 - \exp\left(-\frac{qV_{DS}}{\gamma E(l_0)}\right)\right], \qquad\qquad 8$$

where $I_{DS,sat}$ is the current in saturation regime and $I_{pinch}$ is the maximum current after linear regime. The $\gamma$ parameter is a proportionality factor between the electrostatic energy of interaction and the average transport barrier (randomly oriented surface dipoles lead to a rough energy landscape for charge carriers).

Using Eq. 8 and the calculated interaction energy $E(l_0)$, it is possible to describe the output characteristics for PU-based OLET as a function of $V_{DS}$ for various $V_{GS}$, as shown in Figure 4A solid lines. We found a very good agreement between the experimental data (empty symbols) and the results obtained from Eq. 8 (solid lines), which once again leads us to conclude that the lump behavior is indeed associated to slow polarization centers present in the system.

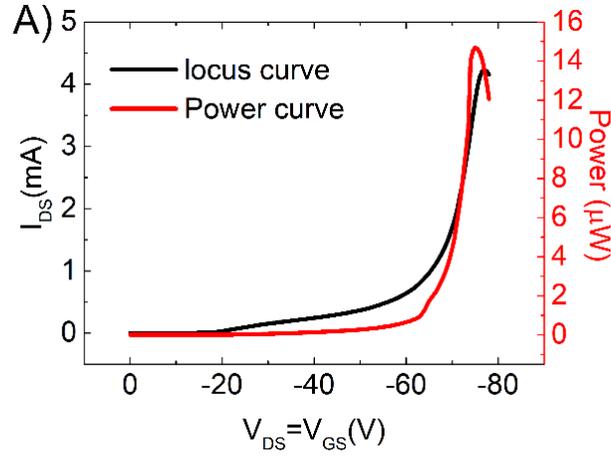

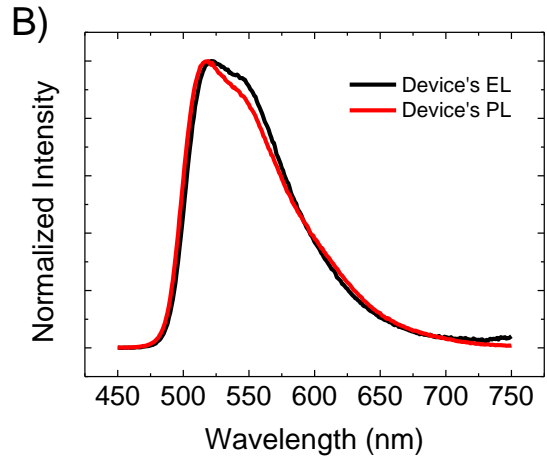

C)

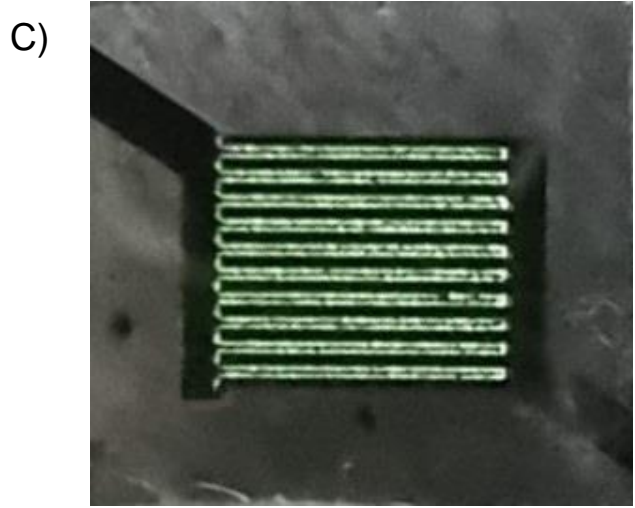

**Figure 5**. A) PU-based OLET's optical power output curves. B) photoluminescence and electroluminescence spectra comparison and C) PU-based OLET under -40V locus mode optical images.

Finally, the optical power associated with the locus sweep is shown in Figure 5. When comparing optical power output, PU-based OLETs also clearly outperforms PMMA-based OLETs. This becomes more evident around -50 V bias condition, where PU surpasses PMMA output under -100 V bias[33]. Also, the maximum EQE% for PU-based devices doubles the EQE% for PMMA-based devices at much lower biases, -60 V and -100 V, respectively. Finally, from Figure 5B it is possible to compare both photoluminescence (PL) and electroluminescence (EL) and conclude that the emission occurs effectively through recombination within the emissive layer. Figure 5C is a device's picture taken in locus mode operation.

## IV.    CONCLUSIONS

PU-based OLETs, besides preserving the possibility of flexible organic electronics via flexible substrates compatibility, exhibit electrical low operating voltages from 0 to -10 V, with threshold voltage of -6.5 V, indeed one of the lowest values reported in literature to date. Such low bias causes less stress to devices by avoiding high electric fields when compared to standard PMMA-based OLETs[43,56].

The high performance obtained by the PU-based OLET results from the low number of traps in the semiconductor/dielectric interface generated by PU's hydrophobic tail. These results suggest that the use of a PU with a larger portion of the flexible and non-polar $-CH_2-$ segment can further improve the performance of PU-based OLET.

We demonstrate that the non-monotonic behavior (lumps) in the $I_{DS}$ vs. $V_{DS}$ curve near the pinch-off regime result from the forced interaction with polarization centers, which increases the interaction energy and consequently reduces mobility. Furthermore, the model used to describe the current in the pinch-off regime is in agreement with the experimental data obtained for PU-based OLETs.

ACKNOWLEDGMENTS

This work was supported and funded by the Brazilian Agencies CAPES, CNPq, FAPERJ, Finep, INCT-Nanomateriais de Carbono and INEO, European Union's Horizon 2020 research and innovation programme and under grant agreement no.780839, (MOLOKO project) grant agreement no. 101016706 (h-ALO project), and italian PNRR MUR project ECS_00000033_ECOSISTER. G.C. gratefully acknowledges FAPERJ Processo E-26/200.627/2022 and E-26/210.391/2022 for financial support. The authors also acknoledge the computational support of Núcleo de Avançado de Computação de Alto Desempenho (NACAD/COPPE/UFRJ), Sistema Nacional de Processamento de Alto Desempenho (SINAPAD) and Centro Nacional de Processamento de Alto Desempenho em São Paulo (CENAPAD-SP).

CONFLICT OF INTEREST

The authors declare that they have no known competing financial interest or personal relationships that could have appeared to influence the work reported in this paper.

AUTHOR CONTRIBUTIONS

A.R.J.B. fabricated the devices, carried out electrical and optical characterization, discussed results and wrote the manuscript. G.C. performed data evaluation and ab initio calculation, discussed results, wrote and revised the manuscript. H.J.C.A. and R.S.C. discussed results and experimental procedure. A.M.S. helped with the fabrication of the devices and the experiments. M.P. performed AFM experiments and discussed results. S.M. performed additional

AFM experiments. E.B. and S.T. suggested OLET's architecture. R.B.C. supervised data evaluation and discussed results. M.M. discussed results and revised the manuscript. M.C conceived the original idea for the manuscript, discussed results, revised the manuscript and led the work.

ABBREVIATIONS

OLETs, Organic Light-Emitting Transistors; OLEDs, Organic Light-Emitting Diodes; OFETs, Organic Field-Effect Transistors; PMMA, poly(methylmethacrylate); rrP3HT, regioregular poly(3-hexylthiophene); PU, polyurethane; C8-BTBT, 2,7-dioctyl[1]benzothienol[3,2-b][1]benzothiophene; TCTA, tris(4-carbazoyl-9-ylphenyl)amine; Ir(ppy)$_3$, fac-tris[2-phenylpyridinato-C2,N]iridium(III); PL, photoluminescence; EL, electroluminescence; AFM, Atomic Force Microscopy; DFT, Density functional theory; SS, subthreshold swing; W, channel width; L, channel length;